\documentclass{siamart171218}

\usepackage{amsfonts}
\usepackage{graphicx}
\usepackage{subfigure}
\usepackage{tikz-qtree}

\usepackage{epstopdf}
\usepackage{algorithmic}
\ifpdf
  \DeclareGraphicsExtensions{.eps,.pdf,.png,.jpg}
\else
  \DeclareGraphicsExtensions{.eps}
\fi

\usepackage{color}
\definecolor{red}{rgb}{1.0,0.,0.}
\definecolor{blue}{rgb}{0.,0.,1.}

\newsiamremark{remark}{Remark}
\newsiamremark{hypothesis}{Hypothesis}
\crefname{hypothesis}{Hypothesis}{Hypotheses}
\newsiamthm{claim}{Claim}

\headers{Shoaling on Steep Continental Slopes}{J. George, D.  I. Ketcheson, and R. J. LeVeque}

\title{Shoaling on Steep Continental Slopes: 
Relating Transmission and Reflection Coefficients to Green's 
Law\thanks{Version of January 13, 2019. Submitted for publication.
\funding{DIK was supported by funding from King Abdullah
University of Science and Technology. 
RJL was supported in part by a subcontract with Science and Technology
Corporation (STC) under NASA Contract NNA10DF26C as part of the
Asteroid Threat Assessment Project (ATAP).}
}}

\author{Jithin George\thanks{Department of Engineering Sciences and Applied
Mathematics, Northwestern University (\email{jithindgeorge93@gmail.com})}
\and David I. Ketcheson\thanks{Computer, Electrical, and Mathematical Sciences \& Engineering Division,
King Abdullah University of Science and Technology, 4700 KAUST, Thuwal
23955, Saudi Arabia. (\email{david.ketcheson@kaust.edu.sa)}}
\and Randall J. LeVeque\thanks{Department of Applied Mathematics, University
of Washington, Seattle, WA 98195-3925. (\email{rjl@uw.edu})}
}

\usepackage{amsopn}

\newcommand{\ignore}[1]{}
\newcommand{\eqn}[1]{(\ref{#1})}
\newcommand{\goto}{\rightarrow}

\newenvironment{mat}{\left[ \begin{array}{ccccccccccccc}}{\end{array}\right]}
\newenvironment{rmat}{\left[ \begin{array}{rrrrrrrrrrrrr}}{\end{array}\right]}
\newcommand\bcm{\begin{mat}}
\newcommand\ecm{\end{mat}}
\newcommand\brm{\begin{rmat}}
\newcommand\erm{\end{rmat}}

\newcommand\CTf{C_T^f}
\newcommand\upeps{^{[\epsilon]}}
\newcommand\upone{^{[1]}}

\ifpdf
\hypersetup{
  pdftitle={An Example Article},
  pdfauthor={D. Doe, P. T. Frank, and J. E. Smith}
}
\fi

\begin{document}

\maketitle

% REQUIRED
\begin{abstract}
The propagation of long waves onto a continental shelf is of great interest in
tsunami modeling and other applications where understanding the amplification
of waves during shoaling is important.
When the linearized shallow water equations are solved with the continental
shelf modeled as a sharp discontinuity, the ratio of the amplitudes
is given by the transmission coefficient.
On the other hand, when the slope is very broad
relative to the wavelength of the incoming wave, then amplification is governed by
Green's Law, which predicts a larger amplification than the
transmission coefficient, and a much smaller amplitude reflection than given by the
reflection coefficient of a sharp interface.
We explore the relation between these results and elucidate the behavior in the
intermediate case of a very steep continental shelf.
\end{abstract}

% REQUIRED

% REQUIRED
%\begin{AMS}

%\end{AMS}

\section{Introduction}
\label{sec:intro}
This paper concerns the shoaling of long waves, such as tsunamis,
as they pass from the deep ocean onto a shallower continental shelf. We seek to
provide a better understanding of the degree to which waves are amplified
during this transition. In particular
we elucidate the connection between the reflection and transmission
that would occur at a sharp interface (a submarine vertical cliff
separating the ocean from the continental shelf) and the prediction
of Green's Law, which holds if the continental slope connecting the
ocean and shelf is instead very gentle, as first derived in \cite{Green1838}.
Green's law predicts a
greater amplification than the transmission coefficient of the sharp
interface, with little reflected energy.  We will show that these two
limiting cases can be naturally connected via a more general solution
that holds for steep continental shelves, with intermediate degrees
of transmission and reflection.

If the wavelength of an ocean wave is significantly greater than
the ocean depth, then the one-dimensional shallow water equations
can be used to model the propagation of a plane wave onto a planar
bathymetry of the sort illustrated in \cref{fig:topo}.  If we also
assume that the amplitude of the wave is very small relative to the
water depth everywhere, then the linearized shallow water equations
can be used as a model.  We consider only this case here.  Green's
Law is often used for shoaling into much shallower water as a wave
approaches a beach, but in this paper we are only concerned with
shoaling onto the continental shelf, which typically has a depth
of a few hundred meters, while the wave of interest has amplitude
at most one or two meters. We also ignore possible effects of
dispersion since our results are primarily of interest for waves that 
are long relative to the width of the continental slope and hence
very long relative to the ocean depth.
This work complements many other discussions in the literature of the
applicability of Green's Law in nonlinear and dispersive cases, e.g.,
\cite{DzvonkovskayaHeronEtAl2014}, \cite{ GrilliSubramanyaEtAl1994},
\cite{HeronDzvonkovskaya2015}, \cite{MadsenFuhrmanEtAl2008}, 
\cite{SynolakisGreenlaw1991}.

The results presented here can be extended to other wave propagation
problems to explore the transmission and reflection of waves passing through
heterogeneous materials.  The linearized shallow water
equations turn out to be a special case in the sense that the ``impedance''
of the material is directly related to the wave speed, whereas for other
wave propagation problems such as acoustics these two parameters describing
the material can vary independently.  We have extended the results presented
in this paper to the more general case in \cite{paper2}, where we also
provide additional mathematical results on the structure of the solution
shown in this paper.  Here we concentrate on providing a better physical
understanding of the process of shoaling on continental shelves in the
intermediate case between Green's Law and reflection/transmission
coefficients.

\subsection{Physical setting}
The linearized shallow water equations in one dimension can be written in
conservation form as
\begin{equation}\label{swe}
\begin{split}
\eta_t(x,t) + (h(x) u(x,t))_x &= 0,\\
u_t + g \eta_x(x,t) &=0,
\end{split}
\end{equation}
where $h(x)$ is the undisturbed fluid depth that we are linearizing about,
$\eta(x,t)$ is the surface elevation (with $\eta = 0$ corresponding to the
undisturbed sea level), $u(x,t)$ is the depth-averaged horizontal velocity,
and $g = 9.81$ m/s$^2$ is the gravitational constant.

\subsection{Transmission and Green's law}
The linear hyperbolic equation \eqn{swe} has the form $q_t(x,t) +
(A(x)q(x,t))_x = 0$, where $q = [\eta, u]$ is the solution vector and
the coefficient matrix $A$ is given by
\begin{equation}\label{Amatrix}
A(x) = \bcm 0& h(x)\\ g&0\ecm.
\end{equation}
This matrix has eigenvalues and corresponding eigenvectors given by
\begin{equation}\label{eigen}
\begin{split}
\lambda_1 &= -\sqrt{g h(x)}, \quad r_1 = \bcm 1 \\ -\sqrt{g / h(x)} \ecm,\\
\lambda_2 &= \sqrt{g h(x)}, \quad r_2 = \bcm 1 \\ \sqrt{g / h(x)} \ecm.
\end{split}
\end{equation}
The eigenvalues are the wave speeds of left- and right-going waves, and the
eigenvectors reveal, for example, that a purely right-going wave on flat
bathymetry with $h(x)\equiv h_0$ must have
$u(x,t) = \sqrt{g/h_0}\eta(x,t)$ for all $x$ and $t$ (see, e.g. \cite{LeVeque2002}).

\begin{figure}
\hfil\includegraphics[width=0.7\textwidth]{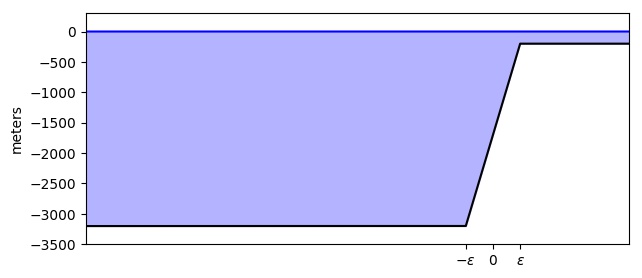}\hfil
\caption{\label{fig:topo}
Bathymetry showing deep ocean for $x<-\epsilon$, a linear continental slope 
for $-\epsilon \leq x \leq \epsilon$, and a continental shelf for
$x>\epsilon$.
  }
\end{figure}

We restrict our attention to piecewise linear bathymetry of the form
\begin{equation}\label{topo}
h(x) = \begin{cases}
h_\ell, & x < -\epsilon,\\
h_\ell + \left(\frac{h_r-h_\ell}{2\epsilon}\right)(x+\epsilon), & -\epsilon \leq x \leq \epsilon,\\
h_r, & x > \epsilon.
\end{cases}
\end{equation}
as shown in \cref{fig:topo}.
Here $h_\ell$ denotes the ocean depth and $h_r$ the depth of the continental
shelf. The region between $-\epsilon$ and $\epsilon$ is known as the continental slope, and
has width $2\epsilon$ with this notation.

Green's Law predicts that the amplitude of a wave moving up a slowly varying
slope will be magnified by a factor of
\begin{equation}\label{CG}
C_G = (h_\ell/h_r)^{1/4}.
\end{equation}
This is valid if the wavelength of the wave is
significantly smaller than the width $2\epsilon$ of the slope.
For the examples, we use $h_\ell = 3200$ m and $h_r = 200$ m, which
are realistic values and were chosen to have a ratio of 16, so that $C_G = 2$.

\begin{figure}
\hfil\includegraphics[width=0.45\textwidth]{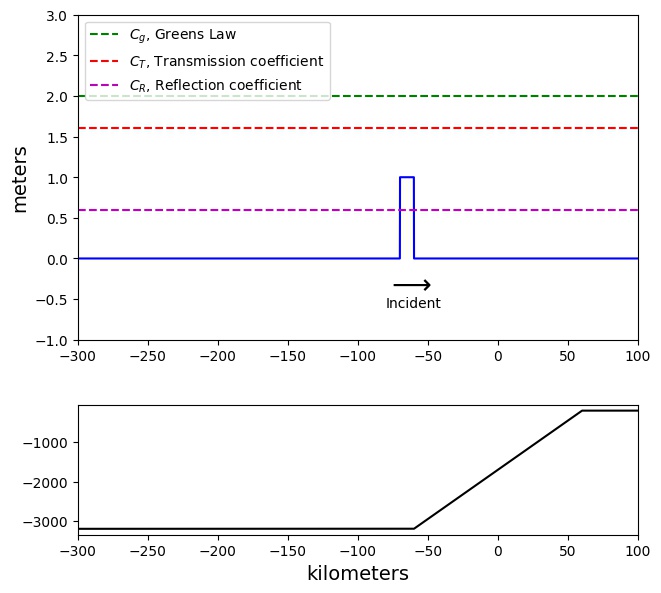}\hfil
\hfil\includegraphics[width=0.45\textwidth]{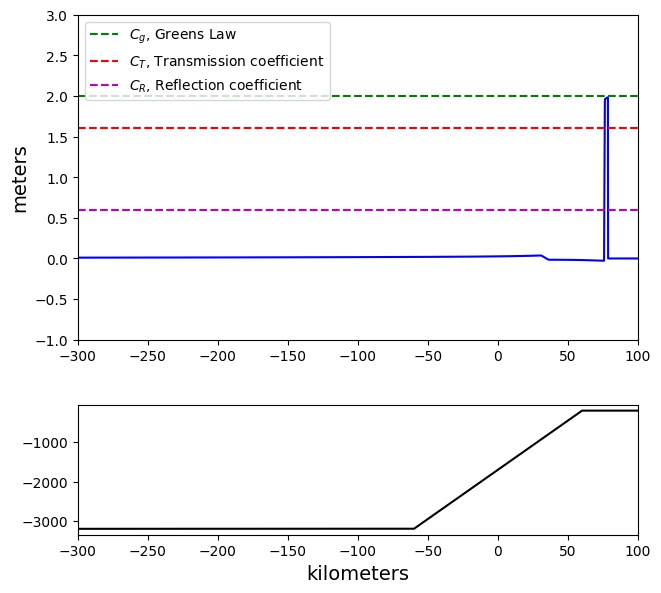}\hfil
\caption{\label{fig:example1a}
Left: The initial data consists of a square pulse of width 10 km and height 1 m
propagating to the right toward a continental slope with $\epsilon = 60$ km.
Right: Solution at a later time $t=1500$ seconds, 
when the pulse has become narrower and
taller on the shelf, with very little reflected energy.  The amplitude is
close to that predicted by Green's Law. The bottom plots show the bathymetry,
with a different vertical scale.
  }

\vskip 10pt

\hfil\includegraphics[width=0.45\textwidth]{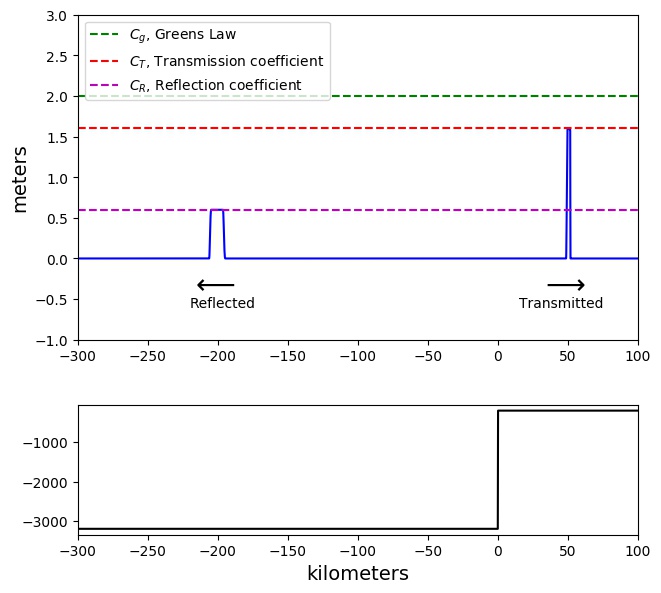}\hfil
\hfil\includegraphics[width=0.45\textwidth]{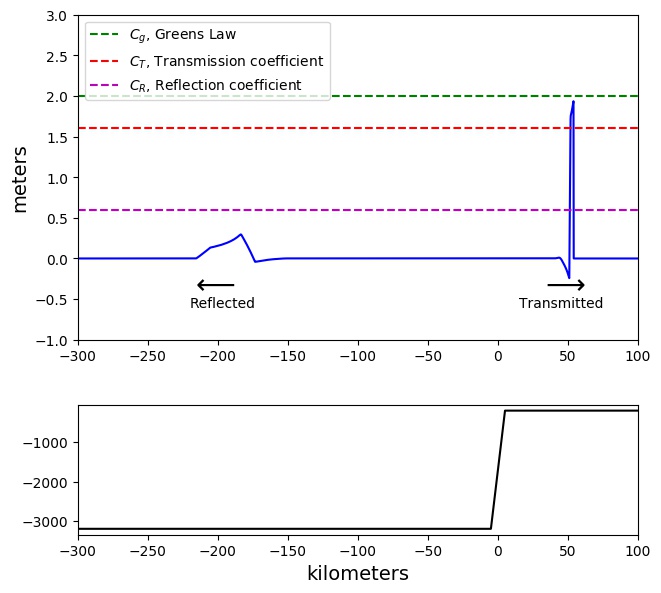}\hfil
\caption{\label{fig:example1b}
Left: Solution when the smooth continental slope of \cref{fig:example1a} has been
replaced with a sharp discontinuity. The solution at $t=1500$ now
consists of a right-going transmitted
pulse on the shelf and a left-going reflected pulse in the ocean.
The initial pulse width is again 10 km.
Right: Solution with a steep slope ($\epsilon = 5$ km), 
showing more reflected energy than in the
case shown in \cref{fig:example1a}.
  }
\end{figure}

To illustrate the shoaling behavior of waves governed by the
shallow water equations, in \cref{fig:example1a,fig:example1b}
we show the solution to these equations when the initial
conditions consist of a square pulse wave approaching a continental slope
with three different choices of the continental slope.  
Although a square pulse is not a realistic tsunami, for the linear
equations over flat bathymetry examining this case is very useful
to help explain the behavior of the solution as the slope is made
more steep. A square pulse can be viewed a pair of discontinuities
(a positive jump followed by a negative jump), and we will first
study the case of a single jump discontinuity propagating toward
the shelf.  We can then use the superposition of two such solutions
to explain the behavior of the square pulse, giving insight also
into the shoaling behavior of a more general wave.

\Cref{fig:example1a} shows a case in which a square pulse moves up a slope
that is broad relative to the width of the pulse, a case where Green's Law
should approximately apply.  The initial surface
displacement is shown on the left, and consists of a pulse with amplitude $A =
1$ m, and with fluid velocity $u = \sqrt{g/h_\ell}$ in the pulse,
to give a purely right-going wave (based on the eigenvectors).  The
plot on the right shows the solution at a later time when the wave
has been amplified due to shoaling by roughly the expected factor
of $C_g = 2$.  The pulse also becomes narrower as the wave speed decreases.
There is a very small amplitude reflected wave in this case, too small to be
seen in the plot.
There is also a bit of a negative tail following the transmitted pulse
that we will say more about later.

\Cref{fig:example1b} shows solutions to the equations with the same initial
conditions as in \cref{fig:example1a} but on steeper slopes.  In the left figure,
the slope is replaced by a jump discontinuity at $x=0$.  In this case Green's
Law does not hold. Instead, the incident pulse is partially transmitted as a
narrower pulse with height $C_TA$ and partially reflected as a left-moving
wave with height $C_RA$.  Using the eigenvectors to the left and right of
$x=0$ and continuity of the solution, we can derive the standard
transmission and reflection coefficients (see \cref{sec:bore}):
\begin{equation}\label{CTR}
C_T = \frac{2\sqrt{h_\ell}}{\sqrt{h_\ell} + \sqrt{h_r}}, \qquad
C_R = \frac{\sqrt{h_\ell} - \sqrt{h_r}}{\sqrt{h_\ell} + \sqrt{h_r}}
= C_T - 1.
\end{equation}
With our choice of $h_\ell$ and $h_r$ we have $C_T = 1.6$ and $C_R = 0.6$.

Finally, the right side of \cref{fig:example1b} shows a case where the width of the
continental slope is comparable to the width of the incoming pulse. In this
case the solution is more complicated, consisting of a transmitted
wave on the shelf that has a peak value close to that predicted by Green's
Law but decreasing behind the peak and has a negative tail, along with a
reflected wave with complicated structure.  As the slope is made more broad, the
amplitude of the reflected wave will decrease and eventually go to zero in the
Green's Law limit.
Our goal is to fully describe waves in this transitional regime of steep
slopes and the manner in which this reflection disappears.
In \cref{sec:implications} we illustrate that realistic tsunamis can fall
into this transitional region, suggesting that a better understanding of
this case is important in practice.

\subsection{Transmission of mass and of energy}
First we note that Green's Law applies to the amplitude of waves, not to their
mass or energy.  Of course the mass of water is conserved and so 
$\int_{-\infty}^\infty (h(x) + \eta(x,t))\,dx$ is constant in time.
Moreover from this or directly from the first conservation law of
\eqn{swe} we see that the ``wave mass''
$\bar \eta \equiv \int_{-\infty}^\infty \eta(x,t)\,dx$
is constant in time and always equal to the initial wave mass. 
Similarly, the second conservation law of \eqn{swe}
shows that $\bar u \equiv \int_{-\infty}^\infty u(x,t)\,dx$ is constant
in time.
Suppose the initial data is identically 0 for $x>-\epsilon$ and consists of some purely
right-going wave with finite mass $\bar\eta$
travelling over the deep ocean.  Then $u(x,0) =
\sqrt{g/h_\ell}\,\eta(x,0)$ for all $x$, and so 
$\bar u = \sqrt{g/h_\ell}\,\bar\eta$.
For large times, the solution
approaches 0 over the continental slope region, and consists of a purely
left-going reflected wave and a purely right-going transmitted wave. Thus for
sufficiently large $t$ we can decompose
$\eta(x,t) = \eta^R(x,t) + \eta^T(x,t)$, and similarly for $u$, where we have
$\eta^R(x,t) \equiv 0$ for $x>-\epsilon$ and
$u^R(x,t) = -\sqrt{g/h_\ell}\,\eta^R(x,t)$ for all $x$, while
$\eta^T(x,t) \equiv 0$ for $x<\epsilon$ and
$u^T(x,t) = \sqrt{g/h_r}\,\eta^T(x,t)$ for all $x$.
If we define $\bar\eta^R = \int_{-\infty}^\infty \eta^R(x,t)\,dx =
\int_{-\infty}^{-\epsilon} \eta(x,t)\, dx$ and
$\bar\eta^T = \int_{-\infty}^\infty \eta^T(x,t)\,dx = \int_{\epsilon}^{\infty}
\eta(x,t)\, dx$, then by conservation of
$\eta$ and $u$, we obtain a linear system of two equations
\begin{equation}\label{TRsys}
\begin{split}
\bar\eta^R + \bar\eta^T &=0,\\
-\sqrt{g/h_\ell}\,\bar\eta^R + \sqrt{g/h_r}\,\bar\eta^T &= \sqrt{g/h_\ell}\,\bar\eta,
\end{split}
\end{equation}
which can be solved to find that the total reflected and transmitted mass are
given respectively by
\begin{equation}\label{etaRT}
\begin{split}
\bar\eta^R &=  \frac{\sqrt{h_\ell} - \sqrt{h_r}}{\sqrt{h_\ell} + \sqrt{h_r}} \eta^0
\equiv \bar C_R \bar\eta,\\
\bar\eta^T &= \frac{2\sqrt{h_r}}{\sqrt{h_\ell} + \sqrt{h_r}}
\equiv \bar C_T \bar\eta.
\end{split}
\end{equation}
Note that $\bar C_T = \sqrt{h_r/h_\ell}\, C_T$.
As the pulse passes over a discontinuous
jump in bathymetry, the amplitude increases by $C_T$ but its width decreases by
$\sqrt{h_r/h_\ell}$ due to the change in wave speed, and so its mass changes by
the product of these, which is $\bar C_T$.
The reflected wave, on the other hand, has the same width as the incident pulse,
and so $\bar C_R = C_R$.

The transmission and reflection coefficients for mass from \eqn{etaRT} apply
regardless of the width of the continental slope, and in fact are the same for
any specified variation of $h(x)$ provided it varies only over a finite region
and is identically equal to $h_\ell$ and $h_r$ away from the slope.  They are
also independent of the shape of the pulse $\eta(x,0)$ as long as the initial
wave is purely right-going, $u(x,0) = \sqrt{g/h_\ell}\eta(x,0)$,
and $\eta(x,0) = 0$ for $x > -\epsilon$.

Note that the square pulse shown in \cref{fig:example1a} for a very broad
continental slope also appears to be transmitted as a square pulse, and the
change in wave speed again suggests the width of the pulse will be reduced by
$\sqrt{h_r/h_\ell}$.  But if the initial amplitude is $A$ then the amplitude of
the transmitted wave is close to $C_G A$ rather than $\bar
C_T A$, and so it seems that too much mass has been transmitted onto the shelf.
However, recall that there is also a small negative trailing wave.  It turns out
this wave has amplitude that decreases as the width of the continental slope
increases, while at the same time it spreads out farther,
in such a way that its total mass is constant and roughly equal to $(\bar C_T -
C_G)A < 0$, which cancels the excess mass that appears in the transmitted pulse.

It is interesting to consider the case of a localized variation in bathymetry,
with $h_r = h_l$ and any variations restricted to $-\epsilon < x < \epsilon$,
e.g., an underwater ridge or sill.
Then $\bar C_R=0$ and $\bar C_T = 1$, so the total mass reflected
is zero while the transmitted mass is equal to the mass of the original wave.
This does not mean, of course, that 
the transmitted wave has the same form as the incident wave, nor that
there are no reflections (only that the integral of the reflected wave vanishes).

It is also interesting to note that the total energy propagated onto the shelf
varies as the width of the slope or the initial pulse are varied.  A wave that is
purely left-going or right-going satisfies equipartition of energy between
potential and kinetic energy (see e.g.\ \cite{mjb-dg-rjl:actanum2011}), 
so it suffices to consider, for example, the
potential energy given by $E(x,t) = \rho g \int_{-\infty}^\infty \eta^2(x,t)\,dx.$
where $\rho$ is the density of water (which is assumed to be constant in this
calculation).  Consider the case of a step function continental slope and an initial
pulse of height $A$ and width $w$, for which $E(x,0) = \rho g wA^2$.
The reflected wave has height $C_RA$ and width $w$,
while the transmitted wave has height $C_TA$ and width $w\sqrt{h_r/h_\ell}$.
So the total potential energy is $\rho g (C_R^2 w  + C_T^2 w\sqrt{h_r/h_\ell})A^2 =
\rho g wA^2$, illustrating conservation of energy.
For a very broad slope, for which Green's Law holds, the energy in the reflected
wave goes to zero (even though its mass is constant, it becomes more spread out
with smaller amplitude as the width of the continental slope increases, and the
energy is quadratic in $\eta$, so this integral vanishes).
The transmitted wave carries all of the energy; it
has height $C_GA$ and width $w\sqrt{h_r/h_\ell}$,
and hence energy $\rho g C_G^2 w\sqrt{h_r/h_\ell}A^2 = \rho g wA^2$.

\section{A single jump discontinuity as the approaching wave}
\label{sec:bore}
\begin{figure}
\hfil\includegraphics[width=0.45\textwidth]{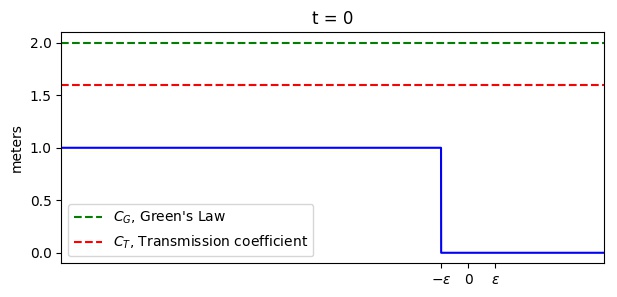}\hfil
\hfil\includegraphics[width=0.45\textwidth]{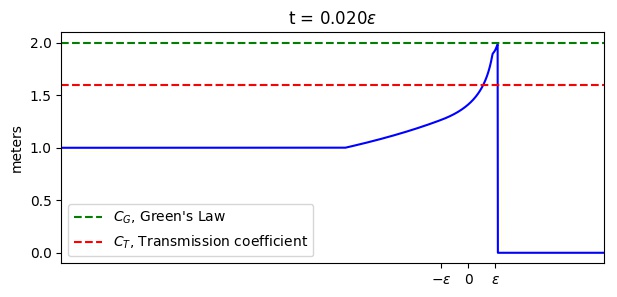}\hfil
\vskip 10pt
\hfil\includegraphics[width=0.45\textwidth]{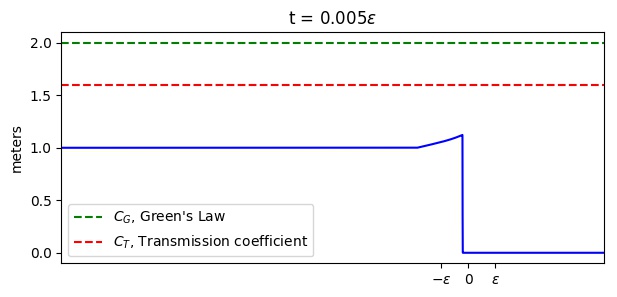}\hfil
\hfil\includegraphics[width=0.45\textwidth]{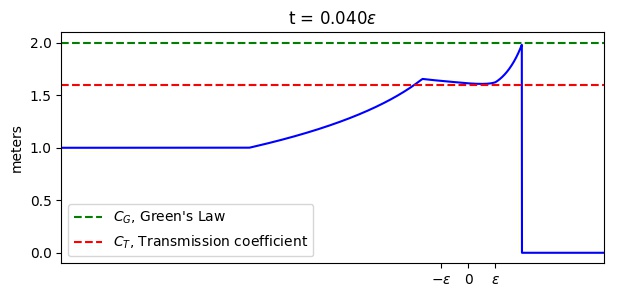}\hfil
\vskip 10pt
\hfil\includegraphics[width=0.45\textwidth]{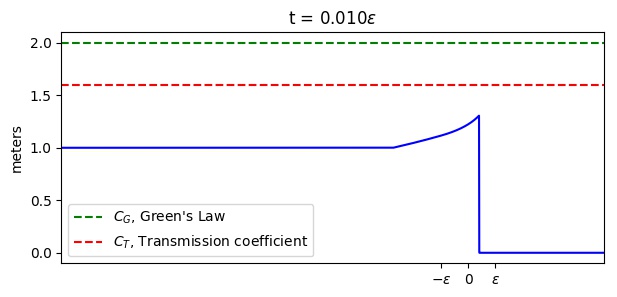}\hfil
\hfil\includegraphics[width=0.45\textwidth]{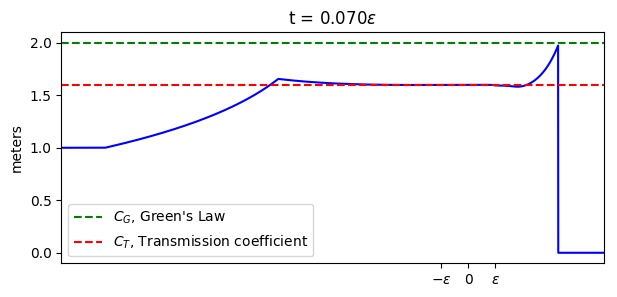}\hfil
\caption{\label{fig:eta1}
Left column: Initial data $\eta\upeps(x,0)$ at $t=0$,
and evolution to $t=0.01\epsilon$, where $\epsilon$ is the
half-width of the continental slope.
Right column: Further evolution to $t =0.07\epsilon$ seconds (when $\epsilon$ is in
meters). At later times the left
going and right-going waves propagate outward over constant bathymetry with
no further change in shape.
  }

\vskip 10pt

\subfigure[]{
\hfil\includegraphics[width=0.45\textwidth]{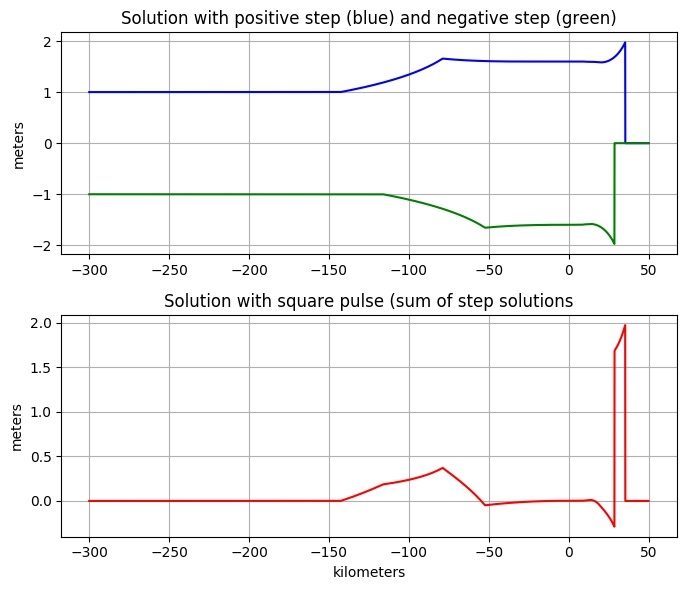}\hfil
}
\subfigure[]{
\hfil\includegraphics[width=0.45\textwidth]{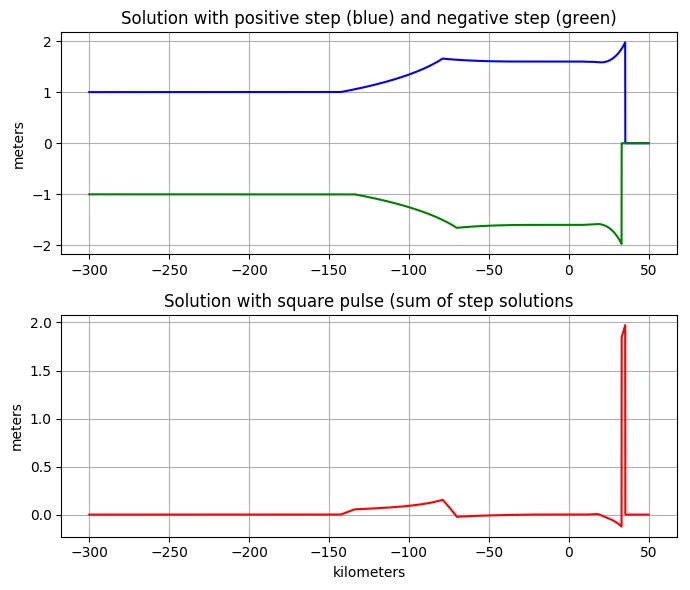}\hfil
}
\caption{\label{fig:step}
The solution to the shoaling problem with square pulse initial data can found as
the linear combination of the single bore solution and a shifted and negated
version of the single bore solution.  (a) A wider pulse has a larger shift between
the two. (b) A narrower pulse has a smaller shift between the two, more
cancellation of the reflected waves, and a transmitted pulse that more closely
resembles the Green's Law prediction.
  }
\end{figure}

To better understand the way in which the waves deform in the intermediate case
between Green's law and pure reflection and transmission,
consider the case of a single jump discontinuity in the initial data 
together with the piecewise linear bathymetry \eqn{topo}.
As data we take
\begin{equation}\label{rpdata}
\eta(x,0) = \begin{cases}
A, & x< -\epsilon,\\
0, & x\geq -\epsilon
\end{cases},
\qquad
u(x,0) = \begin{cases}
\sqrt{g/h_\ell}, & x< -\epsilon,\\
0, & x\geq -\epsilon
\end{cases}.
\end{equation}
This can be viewed as a right-going hydraulic jump (or bore)
moving across the ocean, at
the instant when it first encounters the continental slope,
as shown in the $t=0$ plot in \cref{fig:eta1}.

If $\epsilon=0$ (a jump discontinuity in bathymetry, at the same point as the jump
in the initial data), then the problem has the form of a  ``Riemann
problem''.  Problems of this nature form the basis for much of the theory of
hyperbolic PDEs, and of many numerical methods; see e.g.\
Chap.\ 9 of \cite{LeVeque2002} for more discussion of this 
particular Riemann problem and its solution.
The initial jump in $(\eta,u)$ splits into a single left-going
wave across which $\eta$ jumps from $\eta_\ell$ to some state $\eta_m$,
and a single right-going wave across which $\eta$ jumps from $\eta_m$ to $\eta_r$.
There is a single intermediate state $\eta_m$ since the solution
must be continuous at $x=0$.  Similarly, the
Riemann solution must have a single intermediate state $u_m$.  Moreover, 
due to the form of the eigenvectors displayed in \cref{eigen}, across
the left-going wave we must have $(u_m - u_\ell) = -\sqrt{g/h_\ell}\, (\eta_m -
\eta_\ell)$, while across the right-going wave $(u_r-u_m) = \sqrt{g/h_r}\,
(\eta_r - \eta_m)$.  These conditions result in a system of two equations that,
for the initial data \eqn{rpdata}, yield $\eta_m = C_TA$ and $\eta_m - \eta_\ell =
C_RA$, with $C_T$ and $C_R$ given by \eqn{CTR}.

For any $\epsilon>0$, the solution behaves differently than in the singular case of a
discontinuous continental slope.
As the bore moves up the smooth slope it amplifies and also generates a reflected
wave.  Denote the solution to this problem by $(\eta\upeps(x,t),
u\upeps(x,t))$ to indicate that this solution depends on the parameter $\epsilon$
defining the width of the continental slope.  It is easy to check from the
equations \eqn{swe} that the solution scales with $\epsilon$ in the following
manner: if we calculate the solution $(\eta\upone,u\upone)$
for $\epsilon = 1$ then for any other value  $\epsilon > 0$ we have:
\begin{equation}\label{scalexs}
\eta\upeps(x,t) = \eta\upone(x/\epsilon, t/\epsilon), \qquad
u\upeps(x,t) = u\upone(x/\epsilon, t/\epsilon).
\end{equation}
\Cref{fig:eta1} shows the form of this solution, as computed using Clawpack
\cite{clawpack} on a very fine grid.  Below we will derive the main features
of this solution.

As the bore moves up the continental slope to some point $x>-\epsilon$,
its amplitude increases according to Green's Law  to
$(h(x)/h_\ell)^{1/4}A$, based on the depth $h(x)$ at the point the
bore has reached and the initial depth $h_\ell$ (this is not obvious, 
but will be justified below in section\nobreakspace \ref {sec:R0}).  Upon
reaching
the shelf at $x=\epsilon$, it has
reached an amplitude $C_GA = (h_r/h_\ell)^{1/4}A$, and is followed by a
decrease in amplitude down to the value $C_TA$ given by the transmission
coefficient \eqn{CTR}.  As it propagates up the shelf, it also gives rise to
a left-going wave that increases to
just above $C_TA$ before settling down to the value $C_TA$.  This reflected
wave thus has overall amplitude $C_TA - A = C_RA$.

Below we will explore some details of this solution $\eta\upeps(x,t)$, but first
we note that once we have observed the form of this solution, it is easy to
understand the results shown in \cref{fig:example1a,fig:example1b} and the
transition between them. The square pulse initial data considered in
\cref{sec:intro} can be viewed as a jump discontinuity from $\eta=0$ up to
$\eta=A$ initially located at $x=\epsilon$,
followed by another jump discontinuity from $\eta=A$ back down to
$\eta=0$ initially located at $x = \epsilon - w$, where $w$ is the width of the
square wave.  The solution for each of these initial conditions has the form
discussed above, and the full solution for the pulse is hence (by linearity) the
sum of the solutions for each jump separately.
\Cref{fig:step} shows each of these solutions separately on the top, and the sum
of the two on the bottom. These are shown for two different choices of the width
$w$ of the initial pulse.  As the width decreases, the two waves cancel out
almost everywhere except near the initial peak location, where the amplitude
jumps up to $C_GA$. Hence in the limit of a narrow pulse relative to the width
$\epsilon$ of the continental slope ($w/\epsilon\goto 0$),
we recover the Green's Law limit of a pulse with
amplitude $C_GA$.  For larger values of $w/\epsilon$ there is less cancellation of the
left-going waves and a larger apparent reflected wave.

But note that for any values of $w$ and $\epsilon$, the total mass of the reflected
wave is $wC_RA$, as expected from our earlier discussion.  We can approximate
\begin{equation}\label{deta1}
\begin{split}
\eta(x,t) &= \eta\upone(x/\epsilon, t/\epsilon) - \eta\upone((x-w)/\epsilon, t/\epsilon) \\
& \approx (w/\epsilon) \eta\upone_x(x/\epsilon, t/\epsilon).
\end{split}
\end{equation}
The amplitude vanishes as $w/\epsilon \goto 0$, while
integrating this over the left-going wave shows that the reflected mass
\begin{equation}\label{etaint}
\begin{split}
\int_{-\infty}^{-\epsilon} \eta(x,t)\, dx &\approx (w/\epsilon) \int_{-\infty}^{-\epsilon}
\eta\upone_x(x/\epsilon, t/\epsilon)\, dx\\
& = w(\eta_m - \eta_\ell)\\
& = wC_RA.
\end{split}
\end{equation}
remains constant if $w$ is fixed, independent of $\epsilon$.

\section{Interpretation as a layered medium with piecewise constant bathymetry}
\label{sec:layered}
\begin{figure}
\subfigure[]{
\hfil\includegraphics[width=0.31\textwidth]{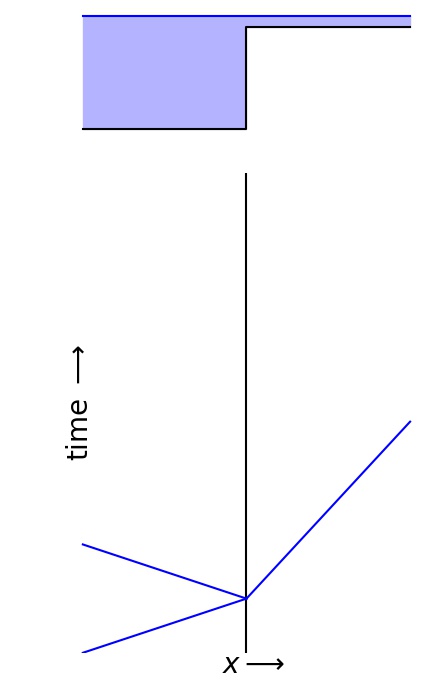}\hfil
\label{fig:layers-a}
}
\subfigure[]{
\hfil\includegraphics[width=0.31\textwidth]{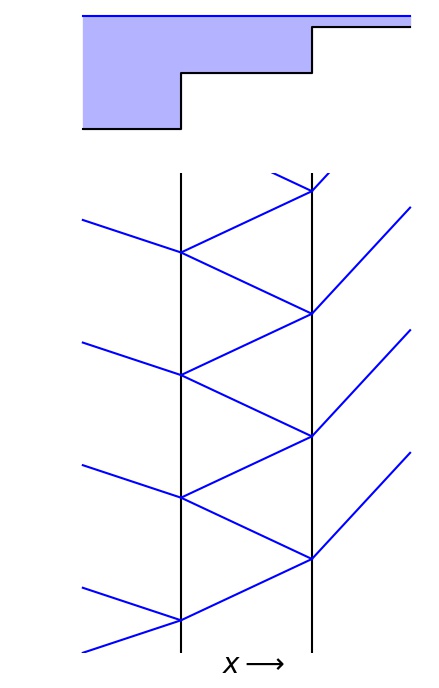}\hfil
\label{fig:layers-b}
}
\subfigure[]{
\hfil\includegraphics[width=0.31\textwidth]{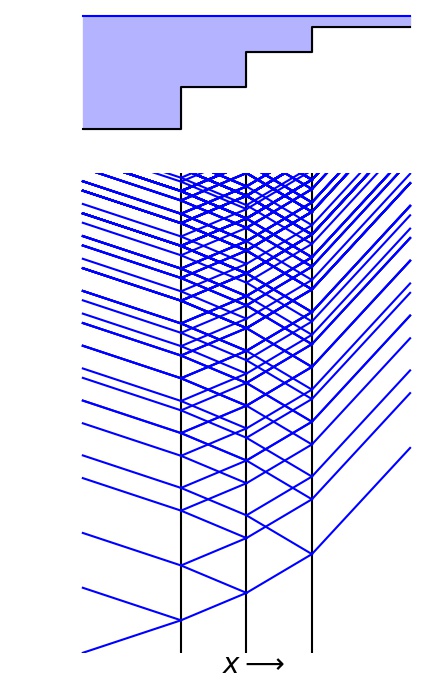}\hfil
\label{fig:layers-c}
}
\caption{\label{fig:topo_and_waves}
Plots in the $x$--$t$ plane showing an initial right-going wave interacting with
a layered medium (i.e.\, bathymetry with jump discontiuities, 
as shown at the top of each plot). 
(a) With a single interface, where only one transmitted and
reflected wave are generated.  (b) With two interfaces and a single intermiate
layer. (c) With three interfaces and two intermediate layers.
  }
\end{figure}

A smooth continental slope can be viewed as the limiting 
case of a sequence of small jump
discontinuities.  Piecewise constant bathymetry of this nature defines a
``layered medium'' in the standard terminology of wave propagation problems;
we provide a more general analysis of this case in \cite{paper2}.  
To understand the behavior of waves through such a medium, it is natural to first
consider a layered medium with only a few intermediate layers.  The single
jump discontinuity between depths $h_\ell$ and $h_r$ that we used to define
the transmission and reflection coefficients in \eqn{CTR} can be partitioned
into $N+1$ discontinuities separating $N$ intermediate layers.
\Cref{fig:topo_and_waves} shows the bathymetry if there are $N=0,~1$ or 2
intermediate layers (top plots) and also shows
how a single initial discontinuity of amplitude
$A$ propagates in the
the $x$--$t$ plane in each case.
With no intermediate layer (\cref{fig:layers-a}, $N=0$), the
wave splits into transmitted and reflected waves only once, with amplitudes
$C_TA$ and $C_RA$ respectively.  With $N=1$ interior layer the solution is
already much more complicated, with internal reflections in the layer that
result in an infinite sequence of waves eventually departing to both the right and to
the left, as shown in \cref{fig:layers-b}.
The amplitude of later waves decays rapidly since each reflection coefficient
is less than one.  We will show in
\cref{sec:one-layer} that the first wave departing to the right  (after
transmission through each layer and no reflection) has amplitude greater than
$C_TA$, but that subsequent departing waves (with 2, 4, 6, or more internal
reflections)  also affect the value of $\eta$ and
asymptotically they sum to $C_TA$.  Similarly, the reflections that depart
from the left (after 1, 3, 5, or more reflections) all sum to $C_RA$.

When the same jump from $h_\ell$ to $h_r$ is split into three discontinuities, as shown
in \cref{fig:layers-c}, then there are many more internal
reflections but the same asymptotic limits still hold, as shown in
\cref{sec:one-layer}.  When additional layers are added, these asymptotic
results continue to hold (see \cite{paper2}).
Moreover, the first transmitted wave exiting to
the right (with no internal reflections) asymptotes to magnitude $C_GA$ as
the number of layers increases, where $C_G$ is the value predicted by Green's
Law.  This can be shown analytically; see \cref{sec:R0}.

In fact, the full wave form of $\eta\upeps(x,t)$ can be worked out
from asymptotic limits of the multilayer case by considering all internal
reflections and transmissions as $N\goto \infty$. However, the number of waves
that must be considered grows exponentially in $N$.
We present this more complicated analysis in \cite{paper2}.  Here we only
discuss two illuminating special cases: a single internal
layer and the first transmitted wave in the limit of infinitely many layers.

\subsection{Single intermediate layer} \label{sec:one-layer}

We derive some results on wave propagation through a layered medium for the
case of a single intermediate layer, as shown in 
in \cref{fig:layers-b} where the intermediate step in bathymetry has
depth $h_m$ with $h_r < h_m < h_\ell$. Define the
following transmission coefficients:
\begin{equation}\label{CTlayers}
C_T^{\ell m} = \frac{2\sqrt{h_\ell}}{\sqrt{h_\ell}+\sqrt{h_m}}, \quad
C_T^{mr} = \frac{2\sqrt{h_m}}{\sqrt{h_m}+\sqrt{h_r}}, \quad
C_T^{m\ell} = \frac{2\sqrt{h_m}}{\sqrt{h_\ell}+\sqrt{h_m}},
\end{equation}
where for example the superscript $mr$ indicates transmission of a right-going wave
from the middle layer to the right layer, and $m\ell$ indicates transmission
of a left-going wave from the middle layer to the left layer.
Similarly define the reflection
coefficients for internal reflections in the intermediate layer as
\begin{equation}\label{CRlayers}
C_R^{mr} =  \frac{\sqrt{h_m} - \sqrt{h_r}}{\sqrt{h_m} + \sqrt{h_r}}, \quad
C_R^{m\ell} =  \frac{\sqrt{h_m} - \sqrt{h_\ell}}{\sqrt{h_m} + \sqrt{h_\ell}}
\end{equation}
Then an initial right-going wave consisting of a discontinuity in $\eta$
of amplitude $A$ leads to a first transmitted wave of amplitude $C_T^{\ell
m}C_T^{mr}A$.  This is larger than $C_TA$, where $C_T$ is the transmission
coefficient with no intermediate layer from \eqn{CTR}, but not as large as $C_GA$,
the value found in \cref{sec:R0} for a smooth transition.   This first wave is
followed by transmitted waves that have undergone $2n$ internal reflections for
$n=1,~2,~3,~\ldots$, each of which has amplitude
\begin{equation}\label{nrefl}
C_T^{\ell m}(C_R^{mr} C_R^{ml})^n C_T^{mr}A.
\end{equation}
Note that $-1 < C_R^{mr} C_R^{ml} < 0$ and so the sum of all the transmitted
waves, including the first wave with no reflections, is given by the geometric series
\begin{equation}\label{tsum}
\sum_{n=0}^\infty C_T^{\ell m}(C_R^{mr} C_R^{ml})^n C_T^{mr}A
= \left(\frac{C_T^{\ell m}C_T^{mr}}{1 - C_R^{mr} C_R^{ml}}\right) A= C_T A
\end{equation}
 As expected, the state behind the right-going waves
decays to the same state $C_TA$ obtained behind the right-going transmitted wave
in the case of no interior layers.

Similarly, the sum of all left-going waves that depart into the deep ocean after
$2n+1$ internal reflections (for $n = 0,~1,~2,~\cdots$) is given by
\begin{equation}\label{rsum}
\sum_{n=0}^\infty C_T^{\ell m}C_R^{mr} (C_R^{ml}C_R^{mr})^n C_T^{m\ell}A
= C_RA,
\end{equation}
where $C_R$ is the reflection coefficient in the  case of no intermediate layers.
But note that there is a time lag between each departing wave, so the
composite reflected wave takes the form of a step function with infinitely
many jumps that decay exponentially with $n$.
%There is a fixed time lag between each departing wave of duration
%$2W_L\sqrt{gh_m}$, where $W_L$ is the width of the layer.

\subsection{Infinitely many layers and the first transmitted wave}\label{sec:R0}

We now consider the amplification of the wave as it is transmitted through each
interface of a layered medium, ignoring all the reflected waves generated in
the process.
The final amplitude of this ``first transmitted wave'' will be denoted by
$\CTf A$, and we will show that $C_T < \CTf < C_G$ and that this transmission
coefficient asymptotes to the Green's Law coefficient
$C_G$ as the number of layers increases.
A different approach to deriving this same result is taken in \cite{paper2}
and \cite{JGeorgeMS}, where more general results are derived for the waves that
also experience internal reflections.

The amplitude of the first transmitted wave in the case of a single intermediate layer
can be represented by $ A_{t_1t_2}$, signifying transmission through the first and second interface: 
\begin{align}
A_{t_1t_2} = C_T^{\ell
m}C_T^{mr}A = \frac{2\sqrt{h_\ell}}{\sqrt{h_{m}}+\sqrt{h_\ell}}
\frac{2\sqrt{h_{m}}}{\sqrt{h_{m}}+\sqrt{h_r}} A.
\end{align}
In the case of two intermediate layers, the amplitude of the first transmitted wave is given by
\begin{align}
A_{t_1t_2t_3} =  \frac{2\sqrt{h_\ell}}{\sqrt{h_{m_1}}+\sqrt{h_\ell}}\frac{2\sqrt{h_{m_1}}}{\sqrt{h_{m_1}}+\sqrt{h_{m_2}}}
\frac{2\sqrt{h_{m_2}}}{\sqrt{h_{m_2}}+\sqrt{h_r}} A.
\end{align}
It can be seen that $A_{t_1t_2t_3}$ is larger than $A_{t_1t_2}$ and  that increasing
the number of layers increases the amplitude of the first transmitted wave.

More generally, consider piecewise constant bathymetry with $N$ intermediate
steps having depths
$h_i$ chosen with $h_i = h_\ell + i\Delta h$ where $\Delta h =
(h_r-h_\ell)/N$ for $i=1,~2,~\ldots,~N$. 
We also define $h_0 = h_\ell$ and $h_{N+1} = h_r$.  Then the
transmission coefficient at the interface between $h_i$ and $h_{i+1}$ is given by
\[
C_{T}^i = \frac{2\sqrt{h_i}}{\sqrt{h_i}+\sqrt{h_i+\Delta h }}
= \frac{2}{1+\sqrt{1+\frac{\Delta h}{h_i}}}.
\]
This allows us to find the continuous limit of the leading amplitude by considering the scenario with infinitely many layers.
The amplitude of the first transmitted wave would then be given by
\begin{align}
\lim_{N\to \infty}\prod_{i=0}^{N} C_{T}^i A =\lim_{N\to \infty}
\prod_{i=0}^{N} \bigg(\frac{2}{1+\sqrt{1+\frac{\Delta h}
{h_i}}}\bigg) A \equiv \CTf A,
\end{align}
Taking the log of $\CTf$ converts the infinite product into a sum,
\begin{equation}
\log(\CTf) 
=\lim_{N\to \infty} \sum_{i=0}^{N}\log\bigg(\frac{2}{1+\sqrt{1+\frac{\Delta h}
{h_i}}}\bigg).
\end{equation}
Using the Taylor expansion then results in
\begin{align}\log(\CTf) &= -\lim_{N\to \infty} \sum_{i=0}^{N}
\bigg(\frac{\Delta h}{4h_i}-\frac{\Delta h^2}{16h_i^2}+\hdots \bigg).
\end{align}
Since the sum of the $\Delta h^2$ and higher
order terms vanish in the limit, we obtain
\begin{align}
\log(\CTf) = -\lim_{N\to \infty} \frac{h_r-h_\ell}{N} \sum_{i=0}^{N}
\frac{1}{4h_i}.
\end{align}
This sum approaches an integral in the limit, giving
\begin{align}
\log(\CTf) = - \int_{h_\ell}^{h_r} \frac{1}{4x} dx = -\frac{1}{4} \log\bigg(\frac{h_r}{h_\ell}\bigg)
\end{align}
and hence
\begin{equation}\label{}
\CTf = (h_\ell/h_r)^{1/4}.
\end{equation}
This is exactly the Green's Law coefficient $C_G$ from \cref{CG}, 
showing that the first transmitted wave has amplitude
$C_GA$ in the limit as the number of internal layers goes to infinity, i.e.\
as the step function bathymetry is smoothed out to a continuous continental
slope.  Note also that this argument is independent of the locations $x_i$
corresponding to each jump discontinuity from $h_i$ to $h_{i+1}$, 
and so the same result holds regardless of whether we discretize a
narrow or wide slope, and regardless of the shape of the slope; it
need not be a discretization of a linear slope of the sort shown
in \cref{fig:topo}.

\section{Implications for real tsunamis}
\label{sec:implications}
Both Green's Law and the more general theory presented here are
based on the one-dimensional shallow water equations and would apply to a real
tsunami only if it were a plane wave approaching a continental shelf
in the normal direction, in a case where the shelf is invariant in
the along-shore direction. A real tsunami is typically a complex
wave train by the time it reaches a shelf, and is not necessarily
approaching normal to the shore.  Moreover, the continental shelf
varies greatly over relatively short spatial scales, with undulations
in width and sometimes deep valleys or outcroppings that will focus or
defocus the approaching waves.  Predicting the amplitude
of the wave hitting the beach from the amplitude seaward of the shelf is
generally not possible with any precision using one-dimensional theory alone,
and simulations in two spatial dimensions with the appropriate shelf
bathymetry and approaching wave train should be used for detailed inundation
predictions.

Nonetheless, one-dimensional theory is still often used to get an estimate of
the amplitudes along a coastline, particularly in cases where many tsunamis
must be simulated and nearshore amplitudes estimated over 
a large region of interest, for example in global
probabilistic hazard assessment, e.g.\ \cite{LoritoSelvaEtAl2015}, 
\cite{LovholtGlimsdalEtAl2012}, \cite{ward_asteroid_2000}.  
Hence we believe it is important for
practitioners to understand the implications of the results presented here.

Comparing this theory with a real tsunami is difficult because each tsunami
and shelf bathymetry is unique.  Here we simply show one example.
We have attempted to identify a case where the leading tsunami wave is
relatively planar and normal to the coast and the width of the continental
slope is comparable to the width of the tsunami, where the results of this
paper imply that amplification should be less than what Green's Law predicts.

\begin{figure}
\hfil
\subfigure[]{
\includegraphics[width=0.70\textwidth]{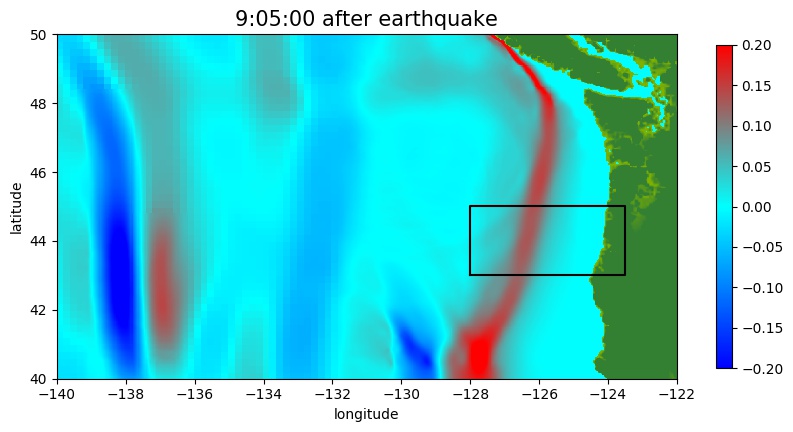}
\label{fig:japan1a}
}
\hfil
%\caption{\label{fig:japan1} 
%  }
%\end{figure}

%\begin{figure}
\hfil
\subfigure[]{
\includegraphics[width=0.7\textwidth]{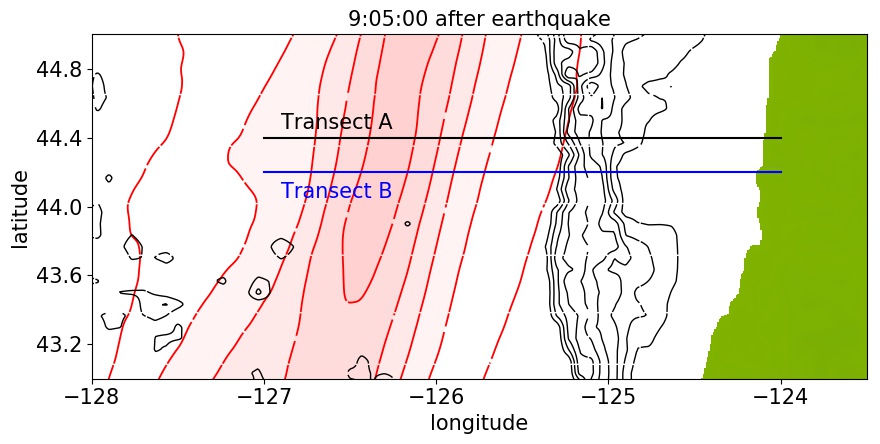}
\label{fig:japan1b}
}

\hfil
\hfil
\subfigure[]{
\includegraphics[width=0.7\textwidth]{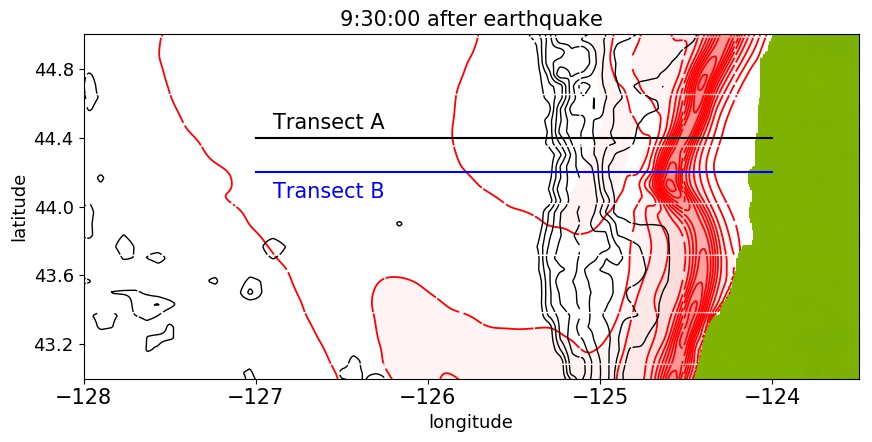}
\label{fig:japan1c}
}
\hfil
\caption{\label{fig:japan1} 
(a) Portion of a simulation of the 2011 Tohoku Tsunami, 9:05 hours after the
earthquake as the leading wave is approaching the west coast of North America.
The black rectangle indicates the region shown in the other two plots.
(b) Zoom on the wave approaching the central Oregon Coast. The black contours show
the topograph of the continental slope (contours at depths of 
$300,~600,~\ldots,~ 2700$ m.
Red contours and shading show surface elevation above sea level, with contours
at $0.025,~0.5,~\ldots,~0.5$ m.
(c) Zoom at a later time 9:30 hours post-quake, after the wave has passed onto the
continental shelf.  Two transects are also indicated, along which
vertical cross-sections of the bathymetry and tsunami are shown in \cref{fig:japan2}.
  }
\end{figure}

\Cref{fig:japan1a} shows a simulation of the tsunami arising from the 2011 
Great East Japan earthquake (also known as the Tohoku event), performed with
the GeoClaw software \cite{clawpack} using an earthquake source model 
provided by the NOAA Center for Tsunami Research and recently used in a
comparison study with tide gauge observations for this event
\cite{GeoClawGauges2017}. This seafloor deformation file along with the
GeoClaw code used to produce these figures is available in the code archive
for this paper \cite{ShoalingCode}.
%\cite{FujiiSatakeEtAl2011}, one of several models that have been developed
%(see e.g.\ \cite{MacInnesGusmanEtAl2013}
%for a comparison of GeoClaw results using several of these with observations).

\Cref{fig:japan1} also shows the leading wave of this tsunami as it is approaching
the coast of Oregon.  Black contours of bathymetry show the continental shelf
and red contours show the wave elevation.  \Cref{fig:japan1b} shows the wave
approaching the shelf, where it is roughly planar with a fairly uniform
amplitude of about 0.15 m. \Cref{fig:japan1c} shows the wave 25 minutes
later, after passing onto the shelf.  The wave has amplified by varying amounts.
The broader shelf around latutude 44.2
causes a folding in of the wave along this latitude (due to the slower wave
speed on the shallow shelf) and amplification along this latitude and lower
amplitudes to the north and particularly to the south (e.g., along latitude 44.0).

\begin{figure}
\hfil\includegraphics[width=0.48\textwidth]{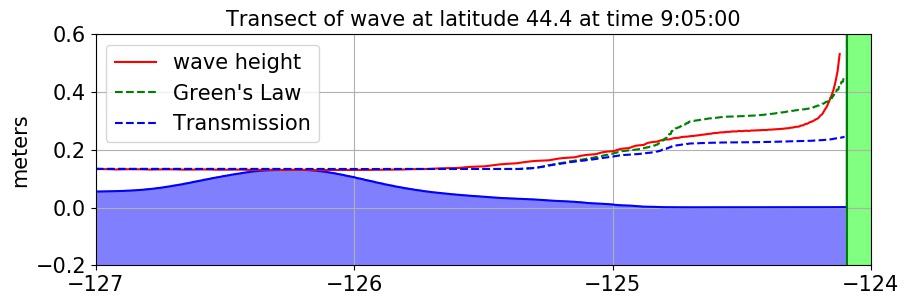}\hfil
\hfil\includegraphics[width=0.48\textwidth]{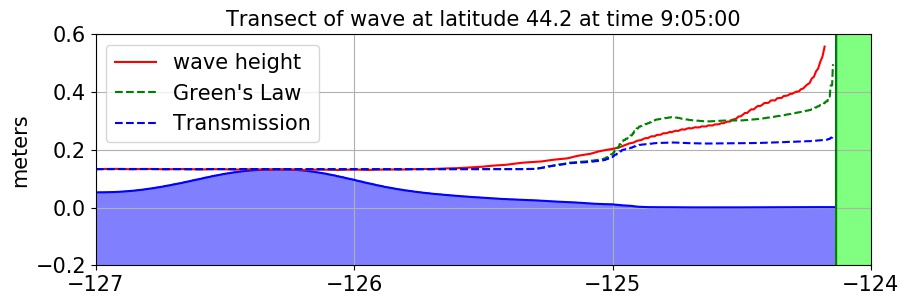}\hfil

\hfil\includegraphics[width=0.48\textwidth]{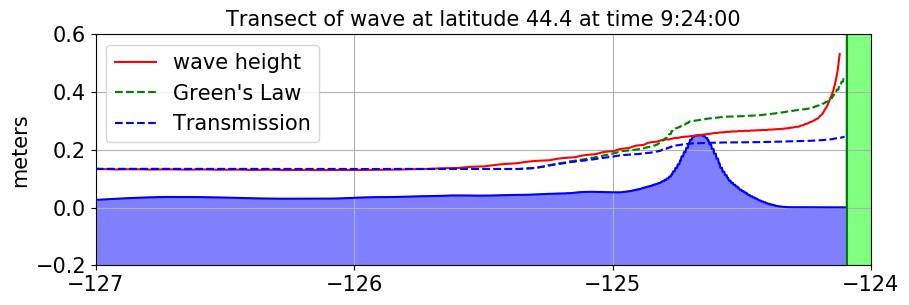}\hfil
\hfil\includegraphics[width=0.48\textwidth]{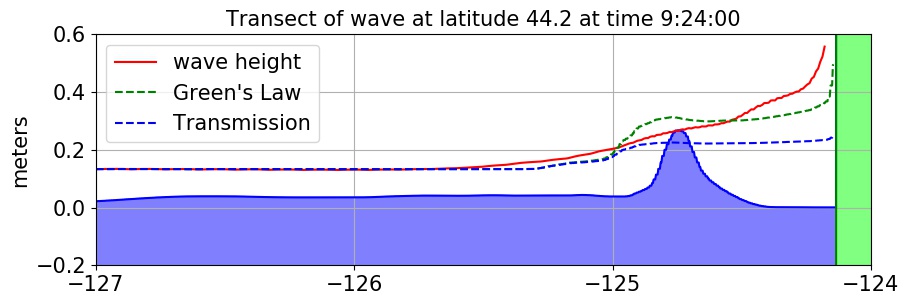}\hfil

\hfil\includegraphics[width=0.48\textwidth]{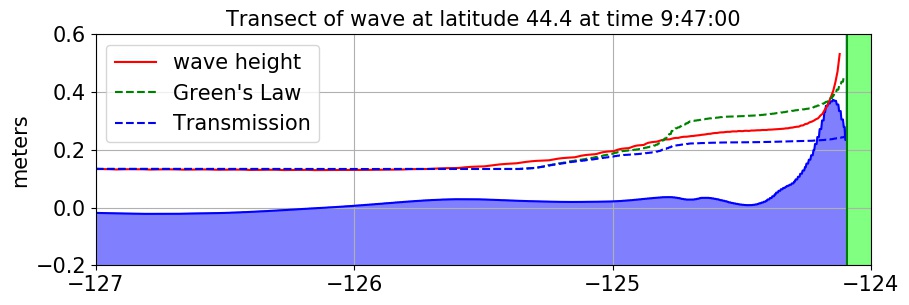}\hfil
\hfil\includegraphics[width=0.48\textwidth]{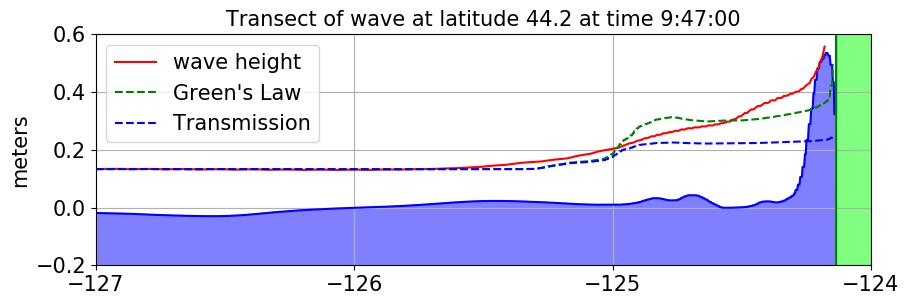}\hfil

\hfil\includegraphics[width=0.48\textwidth]{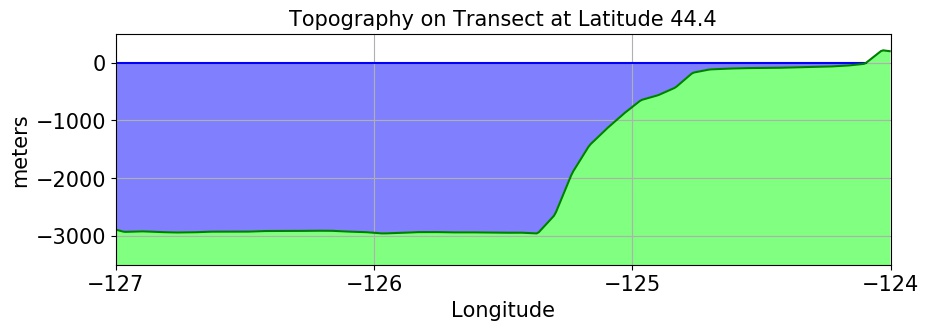}\hfil
\hfil\includegraphics[width=0.48\textwidth]{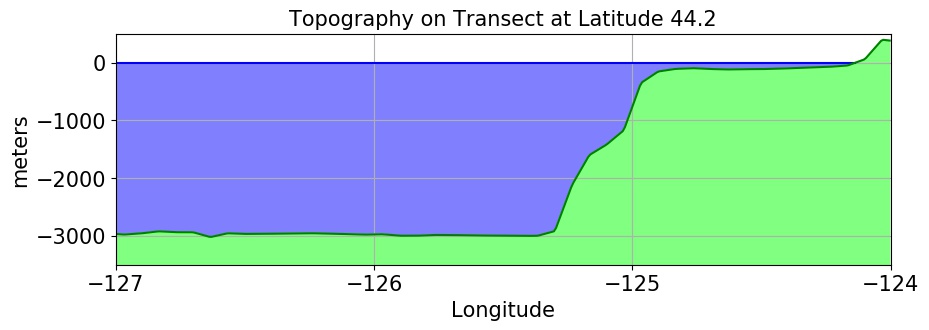}\hfil
\caption{\label{fig:japan2} 
Vertical cross sections of the solution from \cref{fig:japan1} 
(and bathymetry in the bottom plot) along two different transects.  
Left column: Transect A at latitude 44.4$^\circ$N,
Right column: Transect B at latitude 44.2$^\circ$N.
See the text for discussion of the curves shown.}
\end{figure}

In \cref{fig:japan2} we show transects of the bathymetry and of the
solution along two transects, one at latitude 44.4, where the
amplitude is roughly equal to the mean amplitude around this region,
and one at latitude 44.2, where geometric focusing from the shelf
geometry causes greater amplification.  Even in the latter case we
see that the wave amplification as it first passes onto the shelf
is less than what is predicted by Green's Law, which is indicated
by the dashed green curve.  The solution is shown at three times
along each transect, and the red solid line also shows the maximum
sea surface at each point over the full time range from 9:00 to
9:50 hours after the earthquake, showing the amplification as the
wave goes on to the shelf.  The Green's Law value at each point
$x$, based on the depth $h(x)$ at this point relative to the roughly
uniform 3000 m depth seaward of the shelf. The blue dashed line at
each $x$ shows the amplification that would be predicted by the
transmission coefficient from the deep ocean to depth $h(x)$, if
the shelf were replaced by a step discontinuity up to the shelf
depth at this location.

Based on the theory in this paper, we expect the wave to be amplified by more
than the transmission coefficient defined above, but by less than Green's Law
would suggest.  Indeed we see that the red curve of maximum amplitude lies
between the two dashed curves over most of the shelf at latitude 44.4 and at
least initially as the wave moves onto the shelf at latitude 44.2.  
As the wave moves into
much shallower water near the shore, Green's Law roughly applies to predict further
amplification from the shelf depth to the nearshore depth, and we see that the
actual wave amplitude grows even faster near shore due to local bathymetry
effects and runup of the relatively broad wave.  At latitude 44.2 additional
amplification is observed due to the shelf geometry.

\clearpage

\section{Conclusions}
\label{sec:conclusions}

Green's Law is often used to estimate the amplification of tsunamis as they
pass from the deep ocean onto a continental shelf.  This is a good
approximation if the continental slope is sufficiently gentle that the width
of the slope region is large compared to the wavelength.  For steeper slopes
there is less amplification and more reflection of waves, asymptoting to
values given by the transmission and reflection coefficient for a sharp
interface.  Many realistic tsunami applications fall in the intermediate
region, as illustrated for the 2011 Tohoku earthquake tsunami approaching
the Oregon coast in \cref{sec:implications}.  

We have presented a mathematical analysis of the intermediate
behavior that provides new understanding of the connection between
the two limiting cases.  In \cref{sec:bore} we showed that for a
wave consisting of a right-going jump discontinuity approaching a
linear slope, there is a solution that varies in a self-similar
manner as the width of the continental slope is varied. Viewing a
square pulse as a linear superposition of two such waves shows the
manner in which the reflected amplitude and energy decays as the slope width
increases, even though the the reflected mass is independent of
the slope width.  Approximating a continuous slope as a piecewise
constant ``layered medium'', as done in \cref{sec:layered}, shows
that the ``first transmitted wave'' satisfies Green's Law but that
the sum of all transmitted an reflected waves agree with the
transmission and reflection coefficients obtained from a single
sharp interface.  

Additional mathematical analysis that gives a more
complete description of the solution, also in the more general case
of a linear wave equation media for which the wave speed and impedance
can very independently, can be found in our companion paper \cite{paper2}.
The numerical experiments shown in this paper were produced
using the Clawpack software \cite{clawpack}, with code that is available
at \cite{ShoalingCode}.

%\begin{acknowledgements}
\vskip 5pt {\bf Acknowledgements.} 
The authors are grateful to Avi Schwarzschild for stimulating discussions 
in the early phase of this project.
%\end{acknowledgements}

%\bibliographystyle{siamplain}
%\bibliography{../AcousticsPaper/references}

\end{document}